\documentclass[aip,amsmath,amssymb,reprint]{revtex4-1}
\draft
\usepackage{graphicx}  
\usepackage{dcolumn}
\usepackage{amsmath}
\usepackage{makecell}
\usepackage{bm}        
\usepackage{amssymb}   

\usepackage[utf8]{inputenc}
\usepackage[T1]{fontenc}
\usepackage{mathptmx}

\usepackage{float}
\usepackage{color}

\begin{document}

\title{Improving the prediction of glassy dynamics by pinpointing the local cage}


\author{Rinske M. Alkemade}
\affiliation{Soft Condensed Matter, Debye Institute of Nanomaterials Science, Utrecht University, Utrecht, Netherlands }
\author{Frank Smallenburg}
\affiliation{
Universit\'e Paris-Saclay, CNRS, Laboratoire de Physique des Solides, 91405 Orsay, France
}
\author{Laura Filion}
\affiliation{Soft Condensed Matter, Debye Institute of Nanomaterials Science, Utrecht University, Utrecht, Netherlands }
\begin{abstract}
The relationship between structure and dynamics in glassy fluids remains an intriguing open question.  Recent work has shown impressive advances in our ability to predict local dynamics using structural features, most notably due to the use of advanced machine learning techniques.  Here we explore whether a simple linear regression algorithm combined with intelligently chosen structural order parameters can reach the accuracy of the current, most advanced machine learning approaches for predicting dynamic propensity.  To do this we introduce a method to pinpoint the cage state of the initial configuration -- i.e. the configuration consisting of the average particle positions when particle rearrangement is forbidden.  We find that, in comparison to both the initial state and the inherent state,  the structure of the cage state is highly predictive of the long-time dynamics of the system. Moreover, by combining the cage state information with the initial state, we are able to predict dynamic propensities with unprecedentedly high accuracy over a broad regime of time scales, including the caging regime.
\end{abstract}

\maketitle

\section{Introduction}
Understanding the relation between structure and dynamics in glassy systems has sparked extensive discussion over the last few decades\cite{cavagna2009supercooled,berthier2011theoretical,royall2015role,tanaka2019revealing}. While cooling or compressing a glassy system, we typically observe little to no change in structure, while at the same time observing an extreme decrease in dynamics\cite{ediger1996supercooled}. This decrease in dynamics is highly heterogeneous, with increasingly large regions of slow and fast particles as a function of supercooling\cite{ediger1998can, Berthier2020GlassesAA, Candelier2010}. 

One of the approaches to probe the apparent discrepancy between structure and dynamics has been the use of machine learning\cite{Yang-2021, ciarella2022dynamics,cubuk2015identifying, schoenholz2016structural,bapst2020unveiling,
boattini2020autonomously,paret2020assessing,boattini2021averaging,jung2022predicting, shiba2022unraveling,pezzicoli2022se,coslovich2022dimensionality}. By capturing the local structure of particles in terms of parameters and training algorithms to predict the mobility of particles based on these parameters, the idea is that we can learn what aspects of the structure influence the heterogeneous dynamics. 

Over the last two years, the quest for accurate dynamical predictions in glassy systems has led to an explosion of papers introducing new methodologies that compete in predicting the so-called dynamic propensity of simple glassy models. This propensity is defined as the average expected displacement a particle will undergo in a certain time interval when starting from a specific initial configuration \cite{widmer2004reproducible, widmer2007study}. This explosion started in 2020 with the work of Bapst \textit{et al.}\cite{bapst2020unveiling}, where several machine learning methods were trained to predict the dynamic propensity of a Kob-Andersen system\cite{kob1995testing}, with a graph neural network (GNN) performing the best. In 2021, a linear-regression-based algorithm with input parameters that captured structure over several length scales was shown to be able to rival GNNs in predicting the propensity for the same system\cite{boattini2021averaging}. Since then, several works have improved on this feat, by e.g. using physics-informed parameters as input for a deep neural network\cite{jung2022predicting}, by modifying the loss function of a GNN to also consider relative displacements between pairs of particles \cite{shiba2022unraveling}, or by designing GNNs that preserve roto-translation equivariance\cite{pezzicoli2022se}. These works clearly demonstrate that careful consideration of the physics involved can aid in improving the predictive accuracy of these advanced methods.
However, these neural-network based approaches still carry the downside of high complexity. In contrast, due to the simplicity of linear methods, accurately capturing dynamics using linear regression gives a clearer perspective on what structural aspects most strongly drive glassy dynamics. This raises the question: can a clever choice of input parameters boost the performance of linear regression approaches? 

Here, we show that this is indeed the case, and apply linear regression to predict the dynamic propensity of three glass-forming models: hard spheres, harmonic spheres, and the Kob-Andersen model. The main idea of our method is to consider the structure of the system during the caging regime, where each particle is confined by its neighbors in a reasonably well-defined location. By directly incorporating information about this ``cage state'', we show that it is possible to drastically improve the ability of linear regression to predict dynamics, to the point where it even exceeds advanced non-linear machine learning algorithms over a wide range of time scales.

\section{Model, descriptors and prediction method}
\subsection{Model}
Each of our three models consists of two species of particles, labeled $A$ and $B$, with different particle sizes but equal mass $m$. We denote the number of particles of the two species as $N_A$ and $N_B$, such that the total number of particles $N=N_A + N_B$.
Below we discuss each of the models and the statepoints at which we investigate them individually.

\subsubsection{Binary hard-spheres mixture}
The first model we consider consists of hard-sphere particles of two diameters, denoted $\sigma_A$ and $\sigma_B$. The hard-sphere potential for two particles $i$ and $j$ is given by
\begin{equation}
V^\mathrm{HS}(r)= 
\begin{cases}
\infty &\text{for } r\leq \sigma_{ij}\\
0 &\text{for } r > \sigma_{ij},
\end{cases}\end{equation}
where $\sigma_{ij} = (\sigma_i + \sigma_j)/2$. Here, we use a size ratio $\sigma_{B}/\sigma_{A} = 0.85$, and a number ratio $N_A/N = 0.3$. The considered packing fraction of 
$\eta = 0.58$ leads to a structural relaxation time of approximately $\tau_\alpha= 10^4\tau$, with $\tau$ the unit of time given by $\tau = \sqrt{m\sigma_A^2/k_BT}$, $k_B$ Boltzmann's constant and $T$ the temperature. 

\subsubsection{Binary harmonic mixture}
The binary harmonic potential is given by\cite{durian1995foam,berthier2009compressing} 
\begin{equation}
V^\mathrm{Har}(r)= 
\begin{cases}
 \epsilon\left(1-\frac{r}{\sigma_{ij}}\right)^2 &\text{for } r\leq \sigma_{ij}\\
0 &\text{for } r > \sigma_{ij},
\end{cases}\end{equation}
where again $\sigma_{ij} = (\sigma_i + \sigma_j)/2$. Here, we consider the case where $\sigma_B / \sigma_A = 1.4$, $N_A / N = 0.5$. Our state point of interest is at number density $\rho \sigma_A^3 = 0.82$ and temperature $k_BT/\epsilon = 0.0045$, where the structural relaxation time is approximately $\tau_\alpha = 671 \tau$\cite{tah2022fragility}. \\

\subsubsection{Binary Kob-Andersen mixture}
The Kob-Andersen (KA) mixture consists of two particles types \textit{A} and \textit{B} interacting via the Lennard-Jones potential: \cite{kob1995testing}
\begin{equation}V^\mathrm{KA}(r)= 4\epsilon_{ij}\left[\left(\frac{\sigma_{ij}}{r}\right)^{12}-\left(\frac{\sigma_{ij}}{r}\right)^{6}\right],\tag{1}\end{equation}
where $\epsilon_{AA}: \epsilon_{AB}:\epsilon_{BB} = 1: 1.5:0.5$ and $\sigma_{AA}: \sigma_{AB}:\sigma_{BB} = 1: 0.80:0.88$. Note that $(\sigma_{AA}+\sigma_{BB})/2\neq \sigma_{AB}$, i.e. the system is non-additive. The composition of the system is $N_A/N=0.8$. We investigate this system at number density $\rho\sigma_A^3= 1.203$ and temperature  $k_BT/\epsilon_A= 0.44$. The relaxation time of the system is approximately $\tau_\alpha \simeq 3075\tau$\cite{bapst2020unveiling}.


\subsection{Generating initial configurations}

The HS system is simulated using event-driven molecular dynamics\cite{Rapaport2009, smallenburg2022efficient} (EDMD) in the microcanonical ensemble, i.e. at fixed number of particles $N$, volume $V$ and energy $E$. In order to generate snapshots that can serve as initial configurations, we place $N$ particles in the box at a reduced size, and then grow them over time until the desired packing fraction is reached \cite{donev2005neighbor}. Afterwards, the system is equilibrated for $10\tau_\alpha$. 

To simulate both the harmonic and KA systems, we use LAMMPS\cite{LAMMPS}. First we initialize the system by performing a simulation in the canonical ensemble, i.e. at fixed $N$, $V$ and $T$, using a Nose-Hoover thermostat\cite{evans1985nose} at the desired temperature $T$. Afterwards, we equilibrate the system in the microcanonical ensemble,  for $10\tau_\alpha$. 

For each system we equilibrate 100 snapshots, where each snapshot contains 2000 particles. 

\begin{figure}
    \centering
    \includegraphics[width=0.49\textwidth]{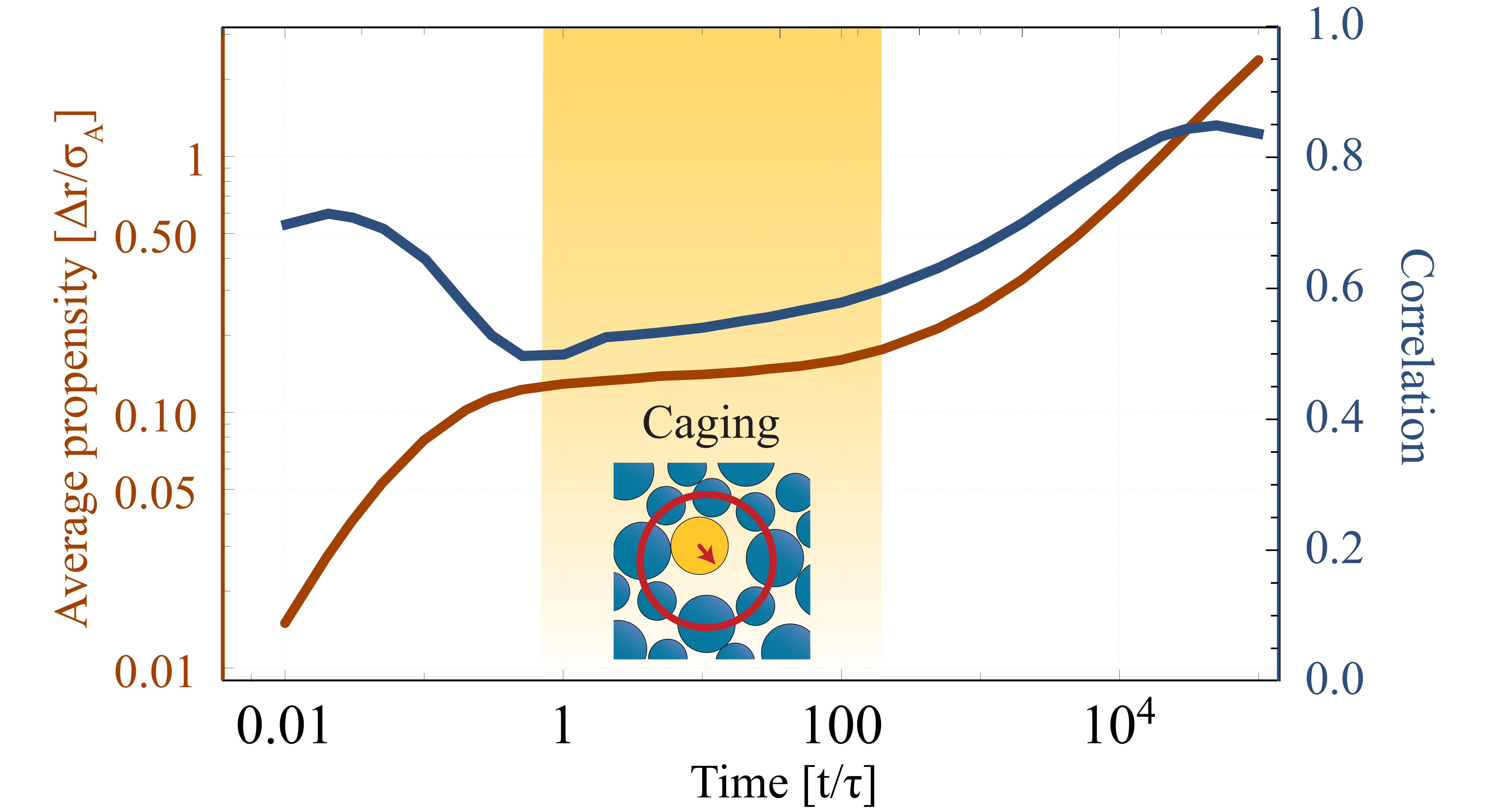}
    \caption{Propensity averaged over all $A$ particles in the system and correlation between the predicted and measured propensity for the same particles, plotted over time for the binary hard sphere mixture at the state point described in this paper. The propensity clearly exhibits three different dynamical regimes: In the ballistic regime, particles have not yet collided with their neighbours, and thus follow a straight ballistic trajectory. In the caging regime, particles move around for an extend period of time in the cage formed by their nearest neighbours. Finally, after particles have escaped the system enters the diffusive regime.  }
    \label{fig:prop}
\end{figure}

\subsection{Dynamic propensity}
As a measure of the dynamical heterogeneity we use the dynamic propensity, which is a quantity that captures the average mobility of particles \cite{widmer2004reproducible, widmer2007study}. To measure it, we simulate the dynamical evolution of each initial configuration 50 times, using distinct initial velocities drawn from a Maxwell-Boltzmann distribution at the desired temperature, i.e. we sample the isoconfigurational ensemble\cite{widmer2004reproducible}. Afterwards the dynamic propensity $\Delta r_i(t)$ of each particle $i$ is obtained by averaging its absolute displacement over the different runs, i.e. 
\begin{equation}
    \Delta r_i(t) = \left\langle |\mathbf{r}_i(t) - \mathbf{r}_i(0)| \right\rangle_\mathrm{conf},\label{eq:propensity}
\end{equation}
where the subscript 'conf' indicates the isoconfigurational average.
We measure the dynamic propensity at logarithmic time intervals between $t/\tau = 0.01$ and $t/\tau = 10\tau_\alpha$. Simulations of the all the three systems are performed in a microcanocical ensemble.

\subsection{Structural Descriptors}
To fit the dynamic propensity, we use standard ridge regression combined with structural order parameters, as previously done in Refs. \onlinecite{boattini2021averaging,alkemade2022comparing}.
The structural parameters include rotationally invariant parameters that capture both the local density as well as the local \textit{n}-fold symmetry. The local density is measured by using radial density functions that capture the density in a shell at distance $r$ and with thickness $2\delta$ from the reference particle. They are defined as 
\begin{equation}
    G^{(0)}_i(r, \delta, s) = \sum_{j\neq i, s_j = s}e^{-\frac{(r_{ij} -r)^2}{2\delta^2}},
    \label{eq:radial}
\end{equation}
where $i$ is the reference particle and $r_{ij}$ is the absolute distance between particles \textit{i} and \textit{j}. The sum goes over all other particles $j$ in the system that are part of particle species \textit{s}. In this paper we include the radial functions up to the fifth minimum in the radial distribution. 

The other structural descriptors we use are based on bond order parameters which express the local structure in terms of spherical harmonics \cite{steinhardt1983bond,lechner2008accurate}.  To compute the parameters we first compute the complex coefficient   
\begin{equation}
    q_i^{(0)}(l, m, r, \delta) = \frac{1}{Z} \sum_{i\neq j} e^{-\frac{(r_{ij} -r)^2}{2\delta^2}}Y^m_l(\mathbf{r}_{ij}). 
    \label{eq:qlm}
\end{equation}
Here $Y^m_l(\mathbf{r}_{ij})$ is the $l^\text{th}$ spherical harmonic function and $m$ is a function that runs from $-l$ to $l$. To normalize the coefficient, we use $Z$, which is given by
\begin{equation}
 Z = \sum_{i\neq j} e^{-\frac{(r_{ij} -r)^2}{2\delta^2}}.
 \end{equation}
Note that due to the exponent, just as with the radial density functions, mainly particles that are within the shell $r-\delta$ to $r+\delta$ contribute to $q_i^{(0)}(l, m, r, \delta)$. The parameters are made rotationally invariant by summing over all possible value of $m$:
 \begin{equation}
    q_i^{(0)}(l, r, \delta) = \sqrt{\frac{4\pi}{2l+1}\sum_{m =-l}^{m =l}|q_i^{(0)}(l, m, r, \delta)|^2}.
    \label{eq:ql}
\end{equation}
Note that the $q_i^{(0)}(l, r, \delta) $ will mainly pick-up the $l$-fold symmetry of the particles structure in each shell. 

In Ref. \onlinecite{boattini2021averaging} it was shown that the prediction of the dynamic propensity via linear regression can be improved when not only the structural parameters of the reference particle itself are included, but also  structural parameters averaged over nearest neighbours. These  averaged structural parameters are obtained via the following recursive formula:
\begin{equation}
    x^{(n)}_i= \frac{\sum_{j: r_{ij}< r_c} x_j^{(n-1)} e^{-r_{ij}/r_c} }{\sum_{j: r_{ij}< r_c} e^{-r_{ij}/r_c}}.
    \label{eq:higherorderbops}
\end{equation}
Here $x_i^{(n)}$ are the structural parameters (which can be both radial density or bond order parameters) of order $n$ for particle \textit{i}. The sum goes over all particles within a certain radius $r_c$ as seen from the reference particle $i$. This $r_c$ is chosen to be located at the second minimum of the radial distribution function, although it was shown in Ref.~\onlinecite{boattini2021averaging} that its exact value has no substantial influence on the order parameters. Boattini \textit{et al} showed including parameters up to three generations significantly improves the predictions. Further information on the descriptors can be found in the Supplementary Information (SI).

\section{Results}

\begin{figure*}[t!]
\begin{tabular}{lclclc}
     & Hard spheres && Harmonic && Kob-Andersen\\
     a) &  & b) & & c) &  \\[0cm]
     & \includegraphics[width=0.3\linewidth]{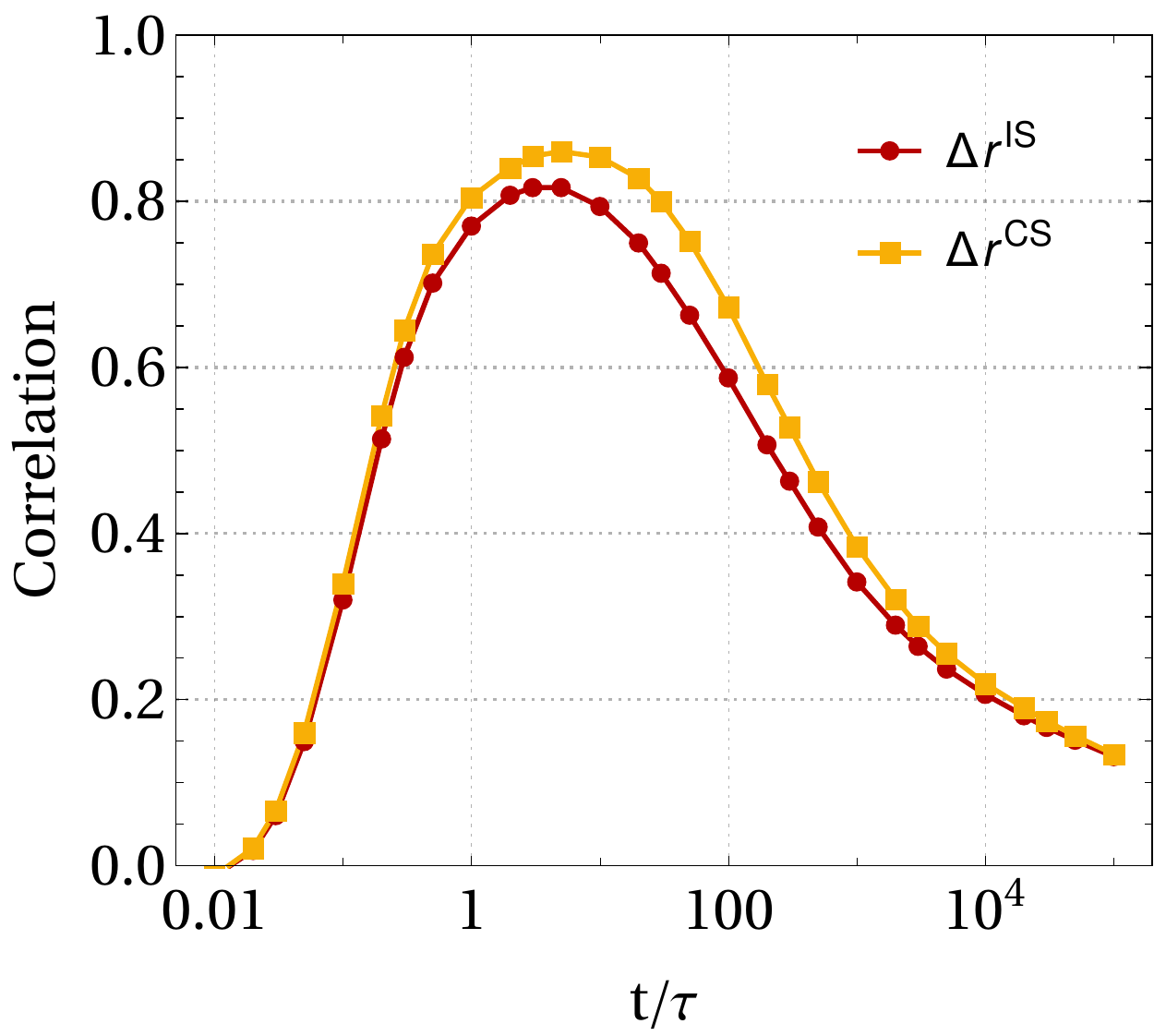} & & \includegraphics[width=0.3\linewidth]{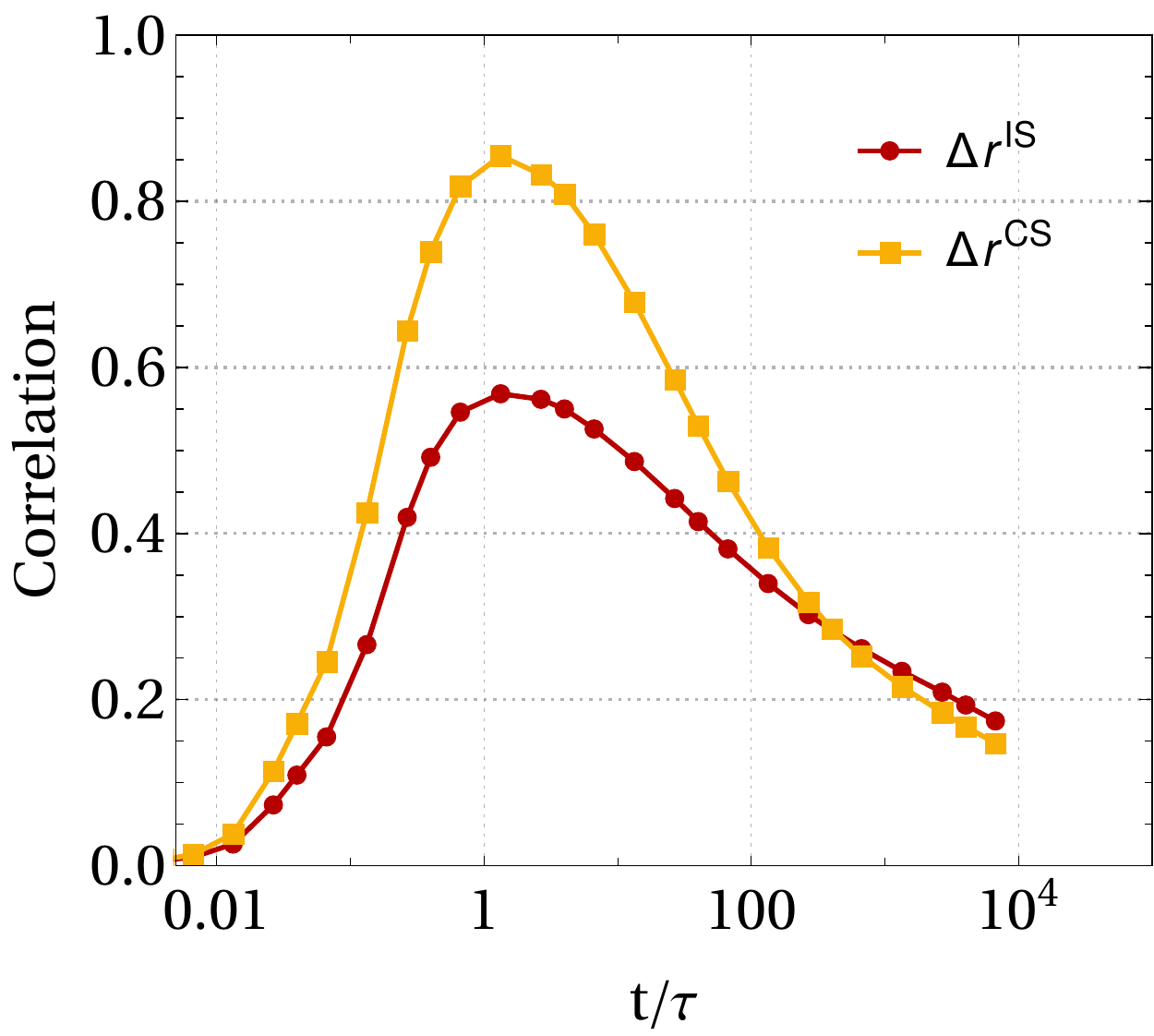}  & &
     \includegraphics[width=0.3\linewidth]{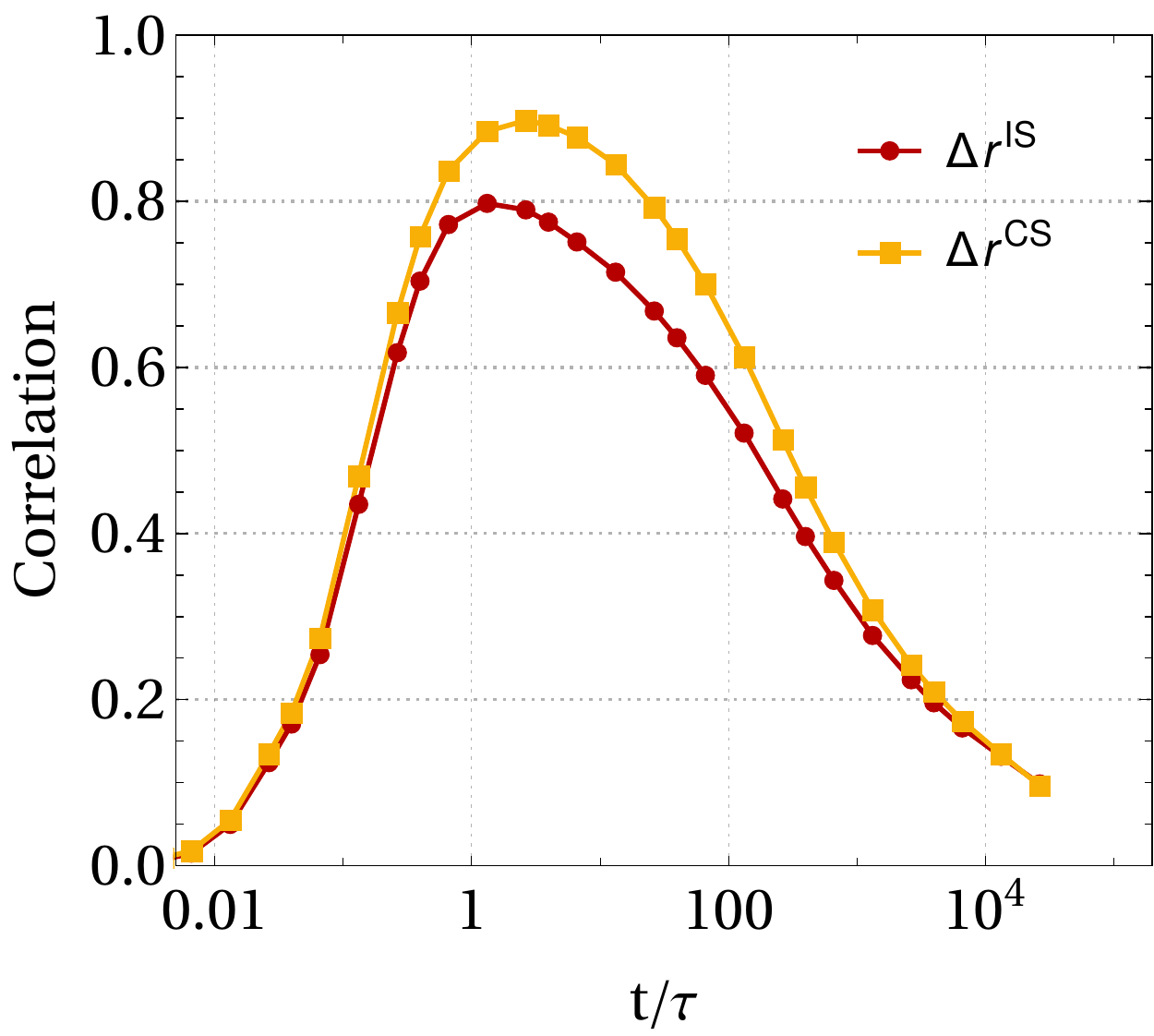} \\
     d) &  & e) & & f) &  \\[0cm]
     & \includegraphics[width=0.3\linewidth]{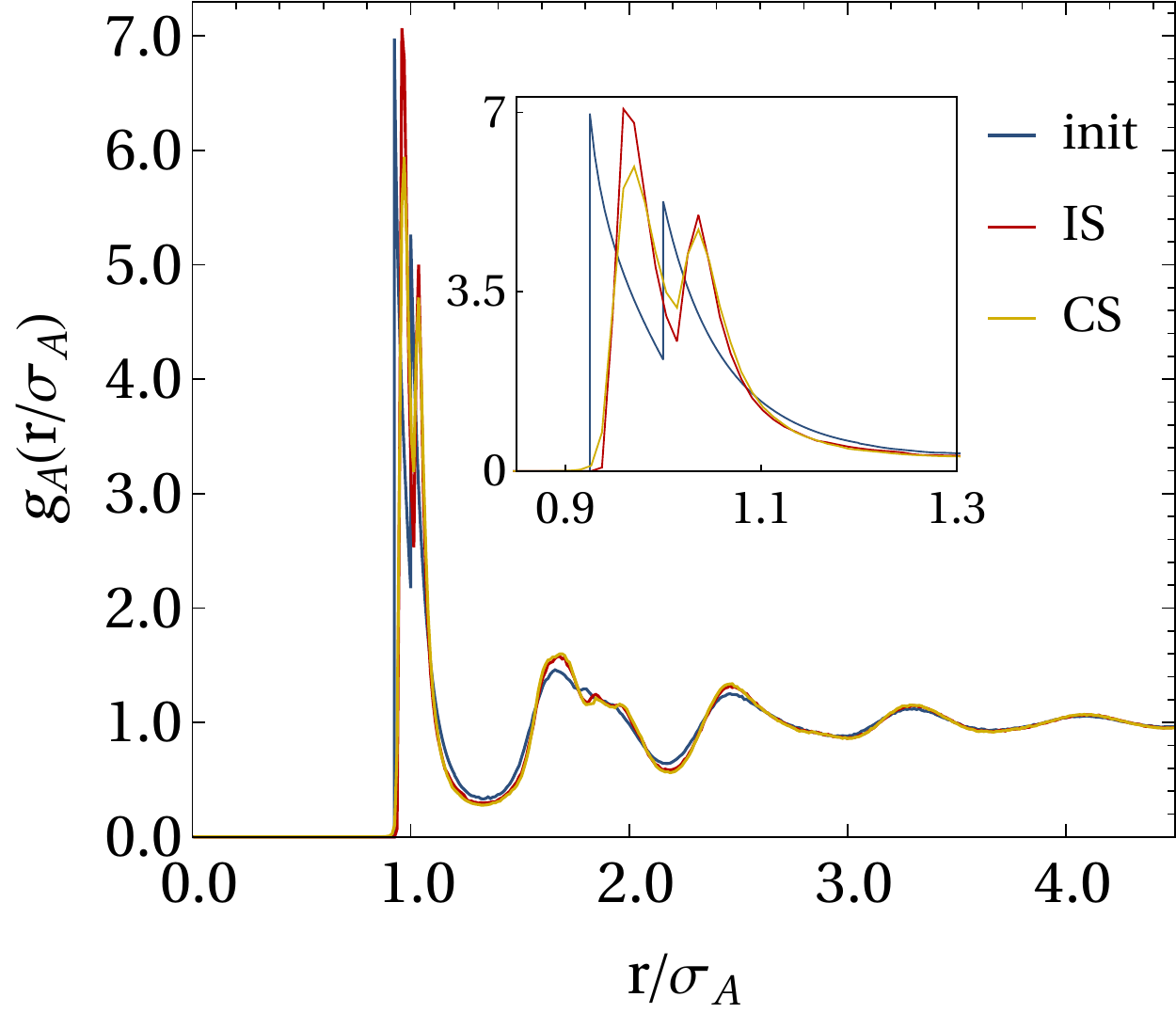} & & \includegraphics[width=0.3\linewidth]{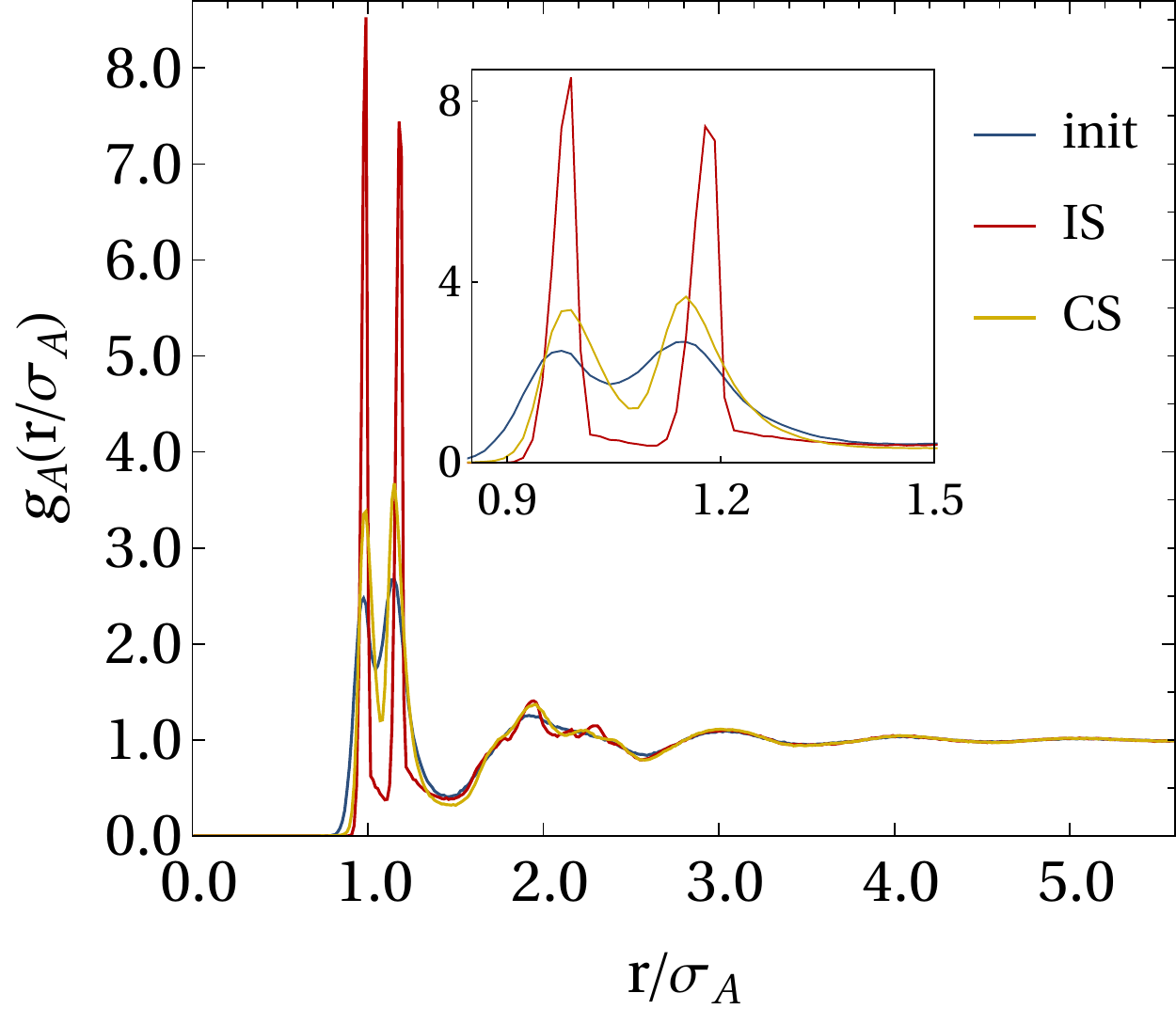}  & &
     \includegraphics[width=0.3\linewidth]{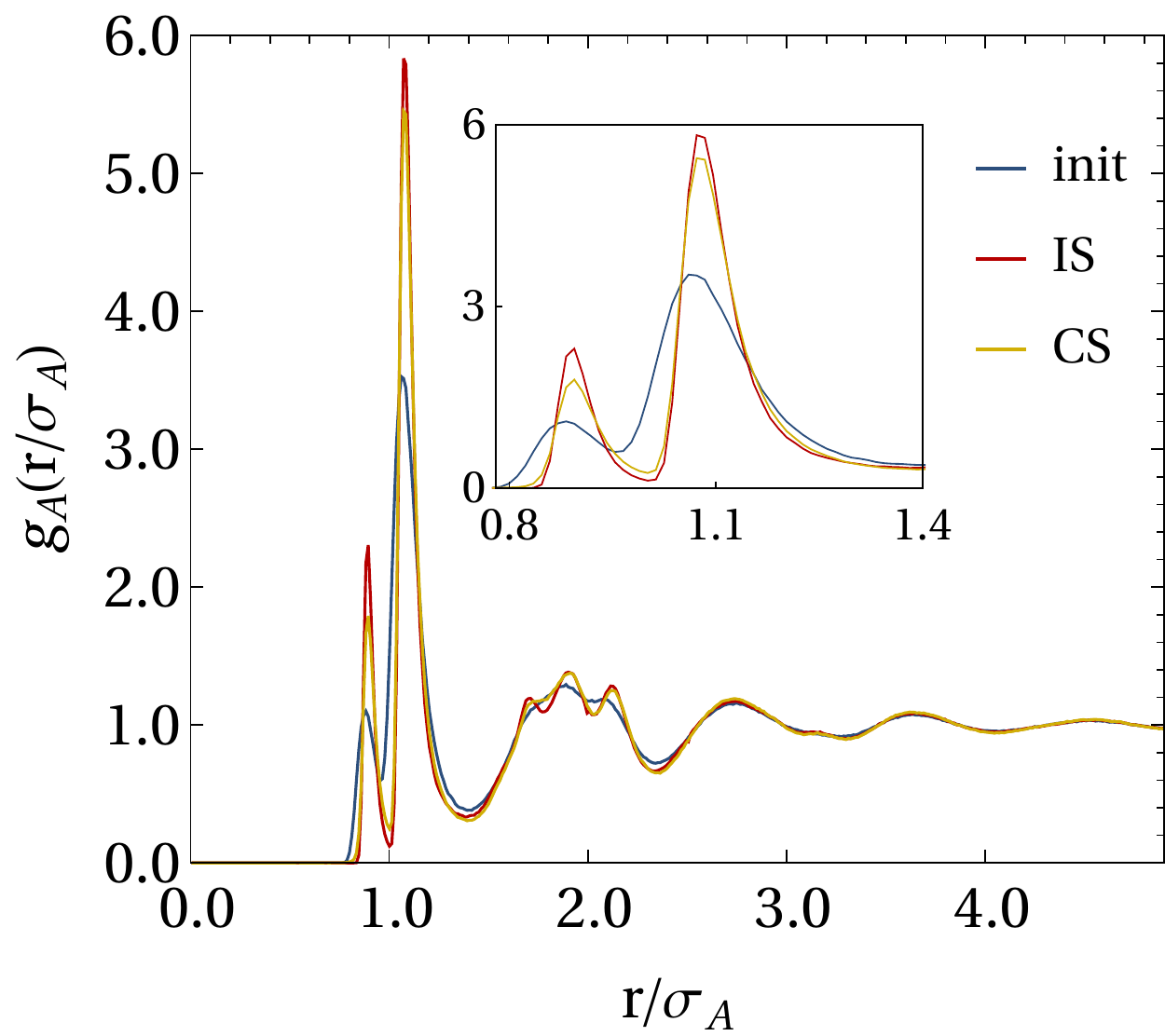} 
\end{tabular}
    \caption[width=1\linewidth]{\textbf{a,b,c)} The Pearson coefficient plotted over time between the propensity and absolute difference between the positions of particles in the initial configuration and the inherent state/cage state ($\Delta r^\text{IS/CS}$) for the $A$-particles in hard spheres (a), harmonic spheres (b), and Kob-Andersen (c). \textbf{d,e,f)}  The total radial distribution function around A-type particles in the initial configuration, the inherent state, and the cage state, for the same three models. }
    \label{fig:cagestates} 
\end{figure*}

A number of recent studies have made significant progress in predicting the dynamic propensity of particles in glassy fluids based on local structural information using a variety of machine learning algorithms \cite{bapst2020unveiling, boattini2021averaging, jung2022predicting, shiba2022unraveling, pezzicoli2022se}. The accuracy of such predictions is typically evaluated using the Pearson correlation coefficient\footnote{The Pearson correlation coefficient is a measure of the degree of linear correlation between two variables. If the Pearson correlation $\mathrm{cor}(\mathbf{X}, \mathbf{Y})$ between two datasets $\mathbf{X}$ and $\mathbf{Y}$ is equal to 1, the two variables are perfectly related via a linear function with positive slope. A value of 0 indicates no linear correlation, while a value of -1 would indicate perfect negative linear correlation.} between the predicted and measured dynamic propensities. A typical example of this correlation as a function of time is shown in Fig. \ref{fig:prop}, where we used linear regression in combination with structural order parameters to predict the propensity in the binary hard-sphere mixture we consider here \cite{alkemade2022comparing}. Interestingly, the correlation is weak in three distinct regimes. Two of these regimes are trivial. The first corresponds to the very short-time regime, where dynamics are dominated by the random choice of initial velocities. Especially for hard spheres, the motion of the particles in this regime is entirely unaffected by local structure, and hence fully unpredictable. The second trivial regime that is hard to predict occurs at long time scales $t \gg \tau_\alpha$, where the system loses memory of its initial configuration. The intriguing regime, where the correlations are weak, lies at intermediate time scales, and corresponds to the times where particles are trapped inside their local cages, as also indicated by the behavior of the average propensity in Fig. \ref{fig:prop}. This lower performance of machine-learned algorithms for predicting propensity in the caging regime is not unique to the results of Fig. \ref{fig:prop}, but has been observed in a variety of studies involving different machine learning methods and different ways to describe the system \cite{jung2022predicting, bapst2020unveiling,boattini2021averaging,alkemade2022comparing}, although recent, more advanced machine learning methods have improved significantly the correlation in this regime \cite{shiba2022unraveling,pezzicoli2022se}. 

Observing the weak correlations in the caging regime, a natural question to ask is then: what structural information do we need to include in order to make accurate predictions about the dynamics of our system in this regime? Since during the caging regime, particles on average move around the so called 'cage center', a rather obvious choice would be to consider the expected average distance between a particle's initial position $\mathbf{r}_i^\mathrm{init}$ and its typical position while in its cage $\mathbf{r}_i^\mathrm{cage}$.  Since the dynamic propensity measures the averaged absolute distance travelled by particles, we would expect this distance between the initial and mean cage position $\Delta r_i^\mathrm{cage} = |\mathbf{r}_i^\mathrm{cage} - \mathbf{r}_i^\mathrm{init}|$ to be a good predictor of the dynamic propensity. While this statement seems logically trivial, an important question is: what is a good estimate of the cage center of a given particle?

A common approach in glass literature for predicting dynamical behavior in a glassy system is to quench the system to its inherent state (IS) (see e.g. \cite{widmer2006predicting, schoenholz2016structural, tong2018revealing, jung2022predicting}). The inherent state of a configuration is defined as the local potential energy minimum that one obtains via a rapid energy minimization\cite{stillinger1982hidden}. Note, however, that this is not exactly the same as the average position of a particle in its cage, as the average should include the effects of thermal fluctuations as well. As such, a natural alternative choice to estimate the position of the cage center is to simply take the average position of each particle, under the constraint that no significant particle rearrangements have happened. We will refer to this second option as the cage state (CS). To explore these two definitions of $\Delta r^\mathrm{cage}$, we calculate for each of our initial configurations both the inherent state and cage state positions. 

For the inherent state, we use the FIRE algorithm proposed by Bitzek \textit{et al.} \cite{bitzek2006structural} to minimize the potential energy. In the case of hard spheres, we apply an effective logarithmic interaction potential proposed by Arceri \textit{et al.}\cite{arceri2020vibrational} in order to obtain an effective inherent state. Specifically, we use the effective interaction potential $V^\mathrm{eff}(r) = -k_BT \log(r-\sigma_{ij})$, with both the forces and interaction energy truncated and shifted to zero at a cutoff radius $r_c = 1.35 \sigma_{ij}$. 

To obtain the cage state,
we use a Monte Carlo (MC) simulation in the canonical ensemble that measures the average positions of each particle $\mathbf{r}_i^{CS} = \langle\mathbf{r}^c_i\rangle$, while restricting the movement of each particle to ensure that it stays inside its initial cage. In order to do this, we reject all MC moves that would move the center of a particle outside of its original Voronoi cell.
Since we consider binary systems, we use an approximate definition of a Voronoi cell which takes into account the particle sizes. In particular, the approximate Voronoi cell for particle $i$ is defined as the collection of points $\mathbf{R}$ for which
\begin{equation}
\frac{\left|\mathbf{R} - \mathbf{r}_i^\mathrm{init} \right|}{\sigma_i} < \frac{\left|\mathbf{R} - \mathbf{r}_j^\mathrm{init} \right|}{\sigma_j}  \forall j \in \mathcal{N}(i),
\end{equation}
where $\mathbf{r}_i^\mathrm{init}$ is the position of particle $i$ in the initial configuration and $\mathcal{N}(i)$ are the nearest neighbours of particle \textit{i} determined by the SANN algorithm \cite{van2012parameter}.
For the KA mixture, we set $\sigma_i = \sigma_{ii}$, since the individual particle sizes are ill-defined.
Note that since the restrictions on the particle positions eliminate the possibility of long-time diffusion, short simulations are sufficient to sample the restricted phase space. Here, we use MC simulations of $5 \cdot 10^5$ initial steps and $10^6$ measuring steps. 

Note that as an alternative to confining each particle to its initial Voronoi cell, we have also explored the possibility of instead confining each particle to a spherical region with a fixed radius $r_c$. When the size of this sphere is chosen to be close to the size of the particle (and hence similar to the size of the Voronoi cell), we find essentially the same results (as shown in the SI). 

\begin{figure*}[t!]
\begin{tabular}{lclclc}
     & Hard spheres && Harmonic && Kob-Andersen\\
     a) &  & b) & & c) &  \\[0cm]
     & \includegraphics[width=0.3\linewidth]{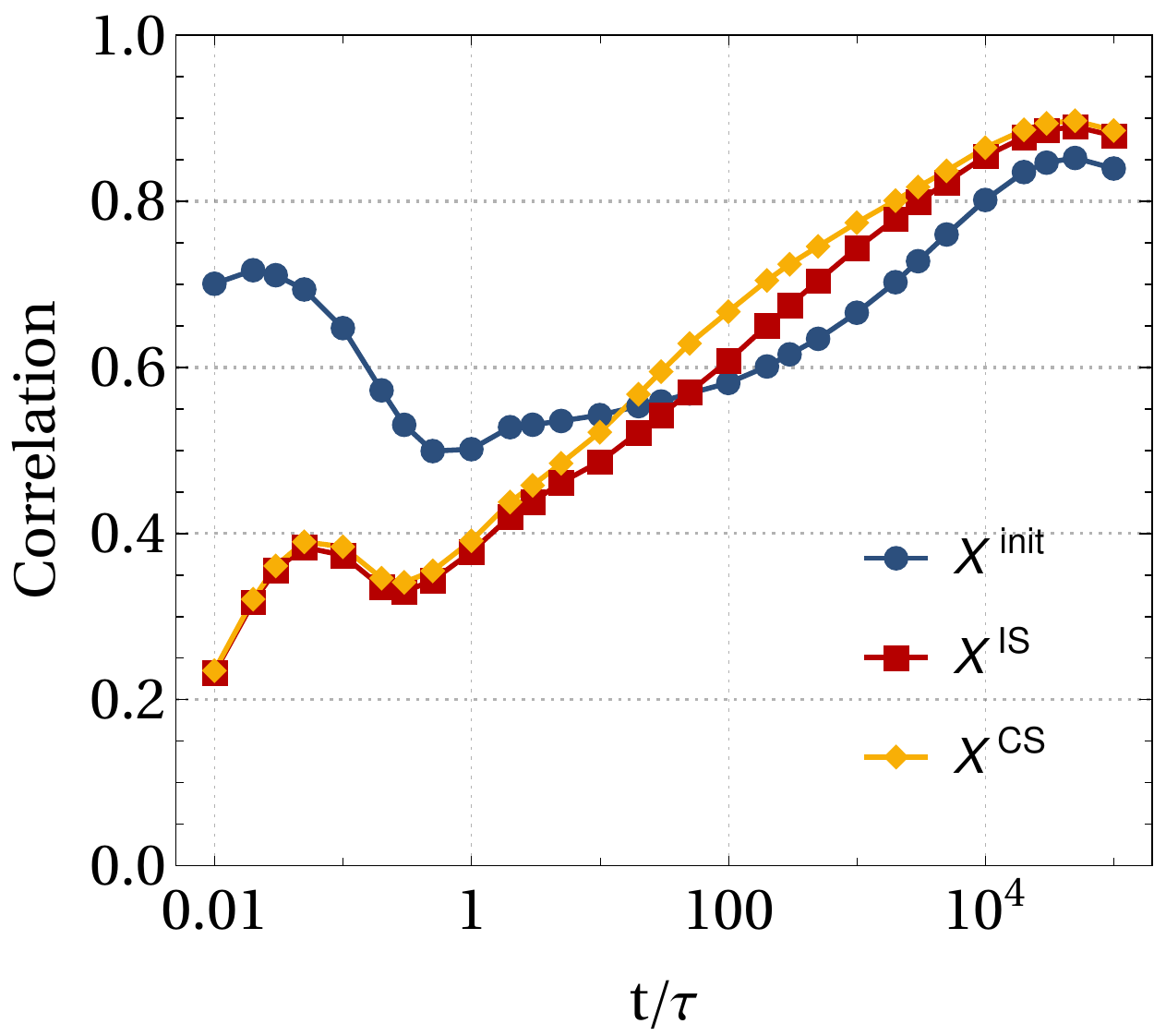} & & \includegraphics[width=0.3\linewidth]{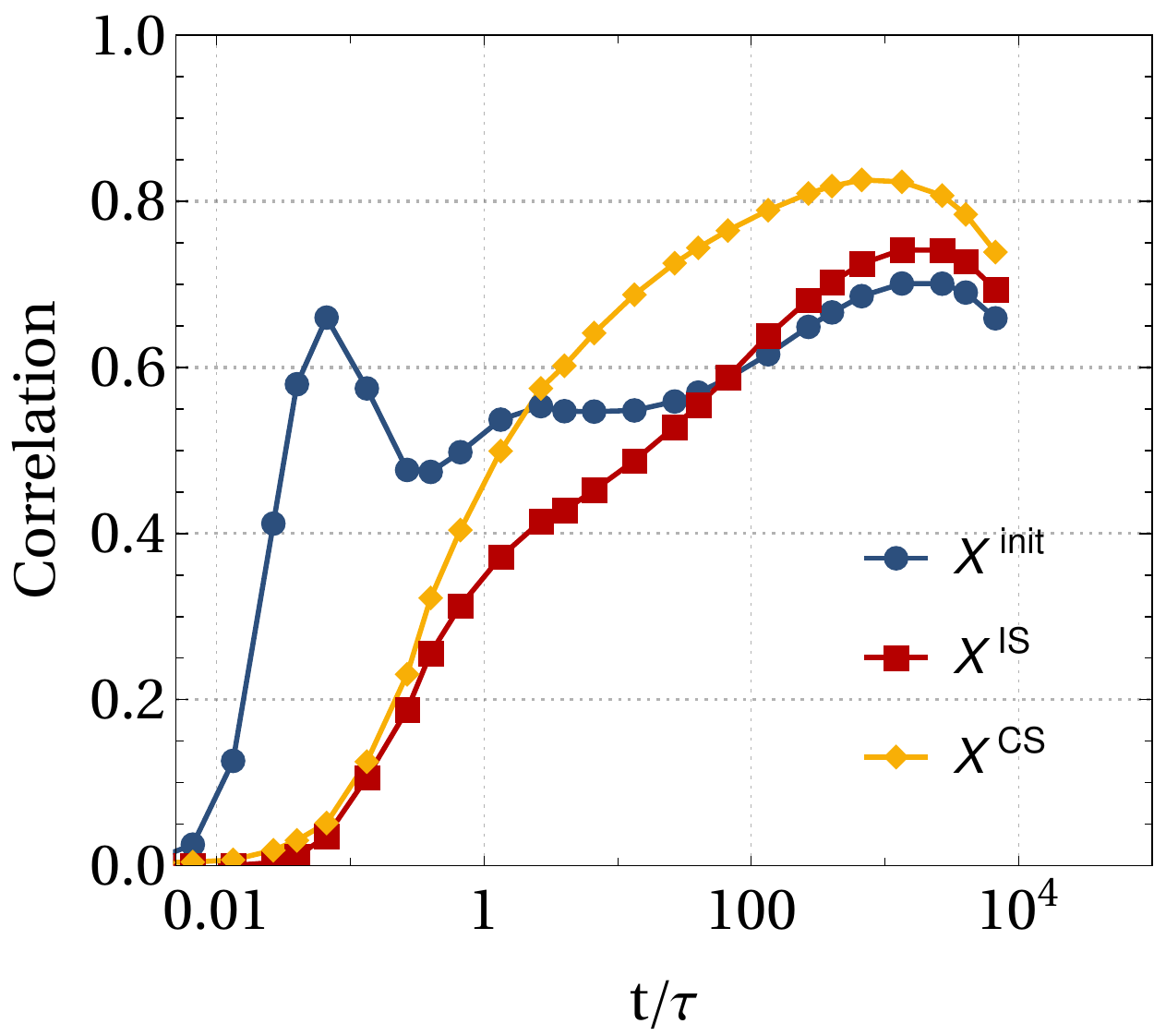}  & &
     \includegraphics[width=0.3\linewidth]{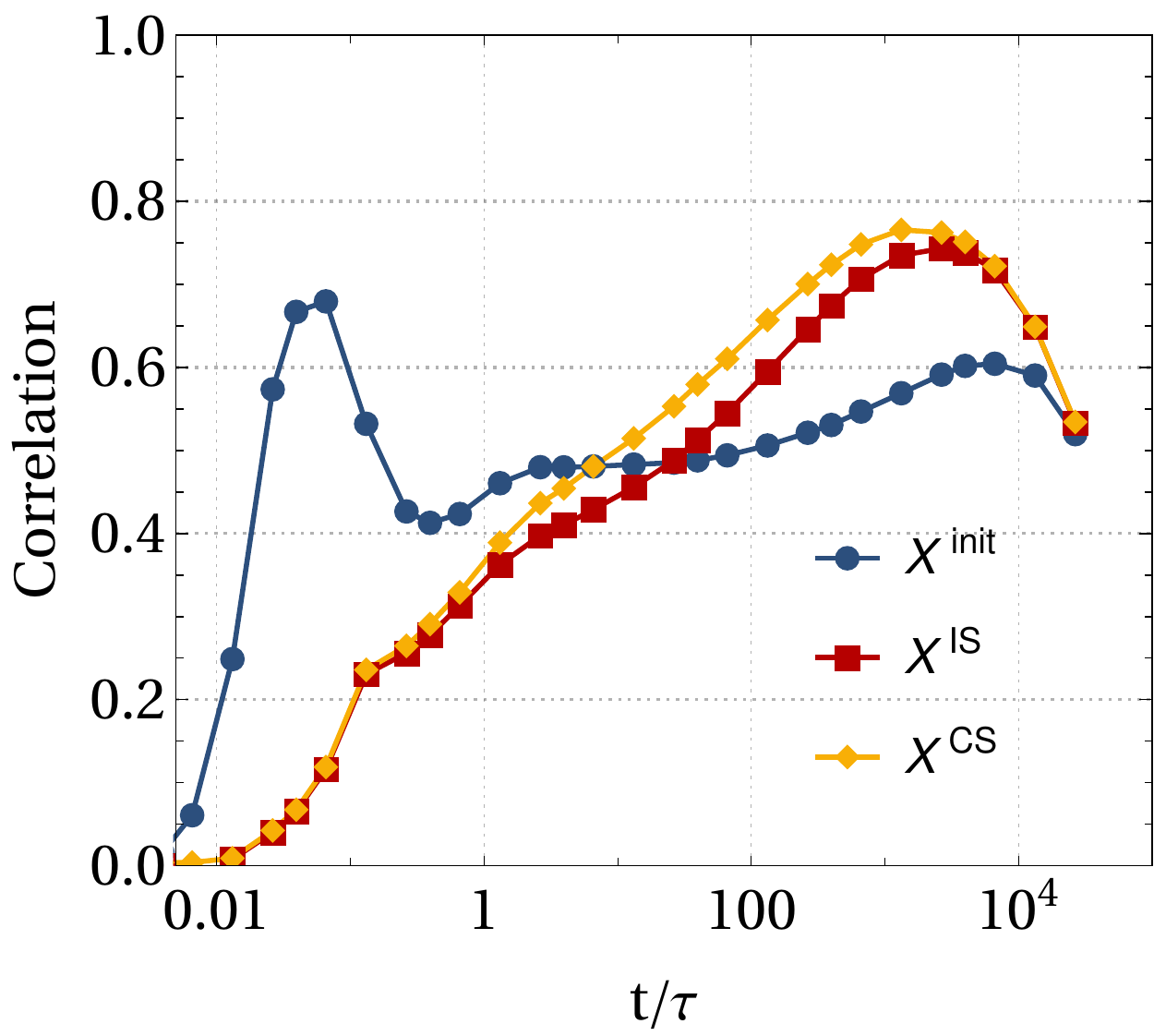} \\
     d) &  & e) & & f) &  \\[0cm]
     & \includegraphics[width=0.3\linewidth]{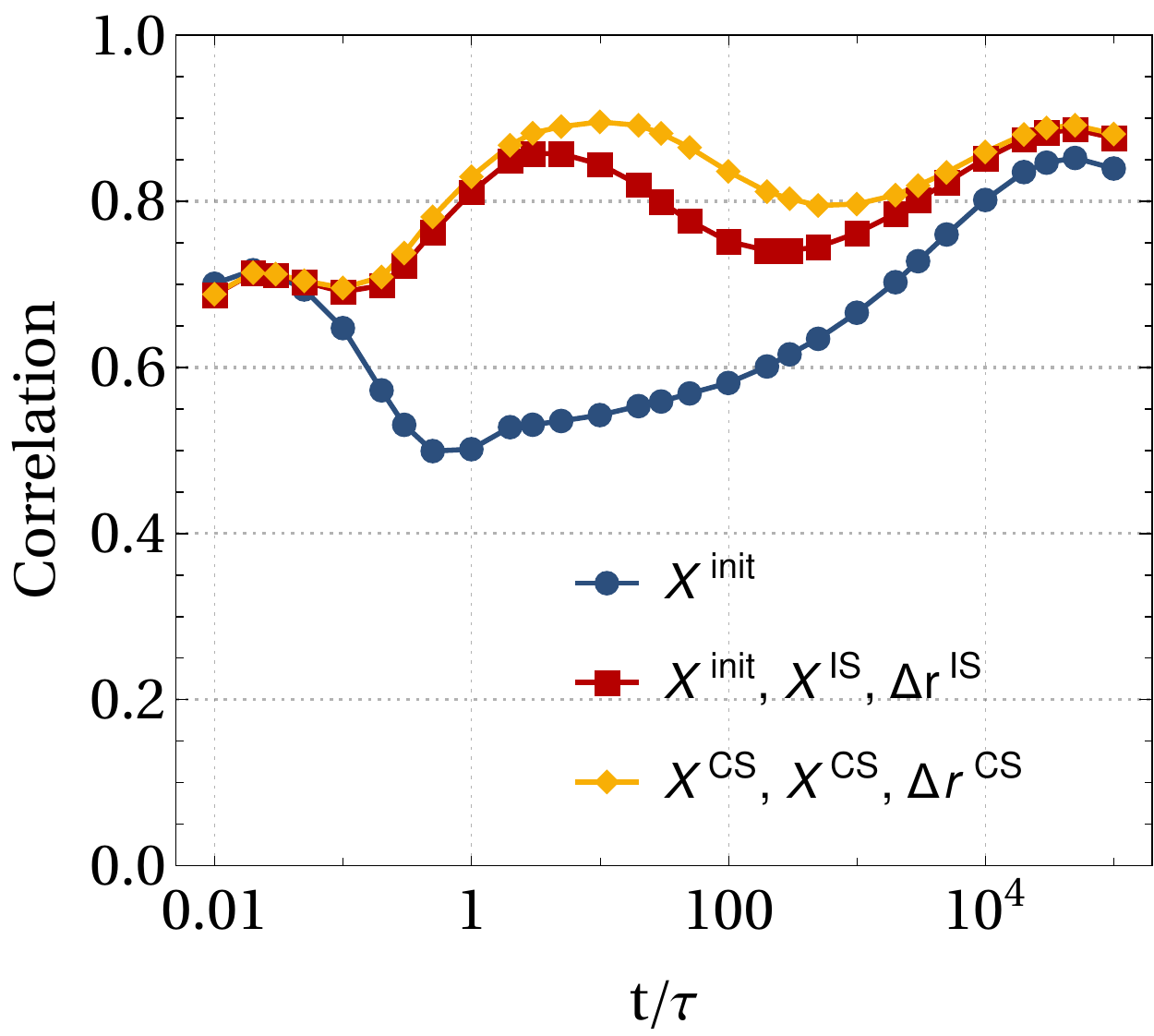} & & \includegraphics[width=0.3\linewidth]{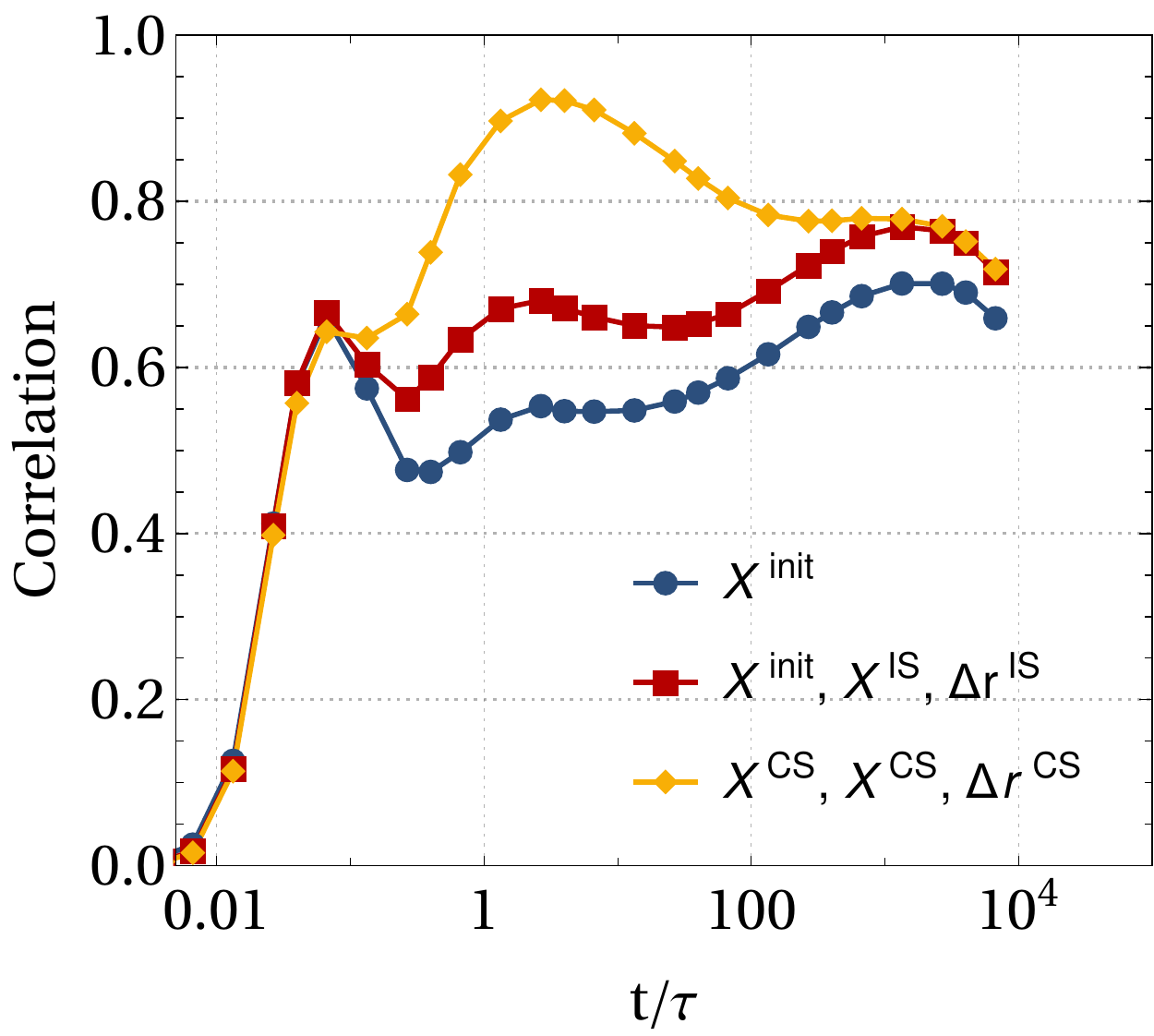}  & &
     \includegraphics[width=0.3\linewidth]{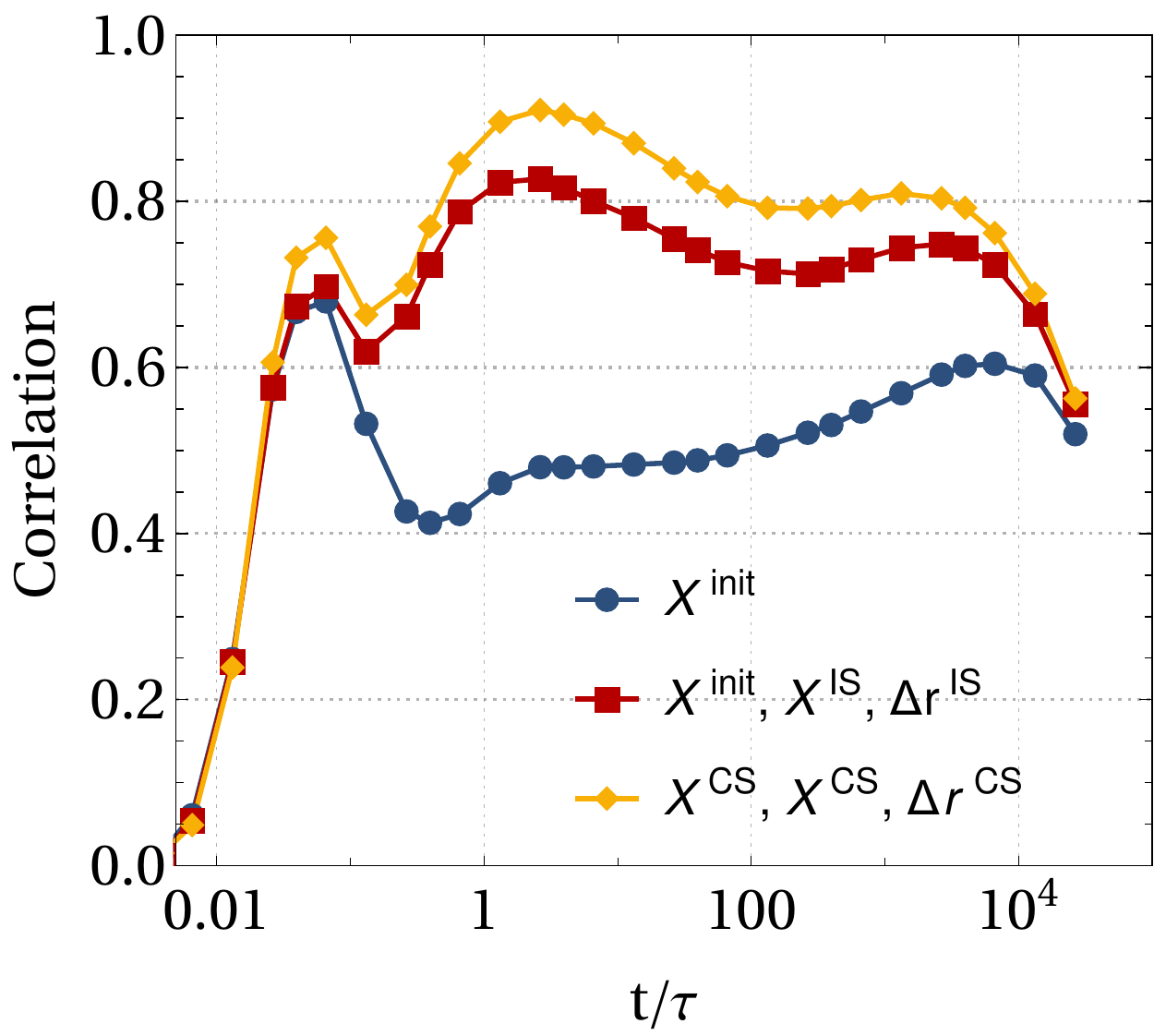} 
\end{tabular}
    \caption[width=1\linewidth]{\textbf{a,b,c)} Pearson correlation coefficient between the dynamic propensity and the prediction of a linear regression model trained on the structural parameters evaluated for the initial-, the inherent- and the cage-state coordinates (i.e. $\mathbf{X}^\text{init}$,  $\mathbf{X}^\text{IS}$ ,$\mathbf{X}^\text{CS}$), for the $A$-particles in hard spheres (a), harmonic spheres (b), and Kob-Andersen (c). \textbf{d,e,f)} Correlation between the measured propensity and the predicted propensity by a linear regression model trained on either only the structural parameters based on the initial positions, or the set of input parameters given by $\{\mathbf{X}^\text{init},\mathbf{X}^\text{IS/CS}, \Delta r^\text{IS/CS} \}$ .}
    \label{fig:correlations} 
\end{figure*}
These two approaches give us two results for $\Delta r^\mathrm{cage}$, which we denote $\Delta r^\mathrm{IS(CS)}$ for the inherent (cage) state. To compare how effective these measurements are at capturing the actual cage center, we plot in Fig. \ref{fig:cagestates}abc the Pearson correlation between the dynamic propensity and both definitions of $\Delta r^\mathrm{cage}$, for all three glass model systems. Note that since here we consider only a single parameter at a time, no linear regression is required to obtain a correlation. As expected, both the inherent state and the cage state provide significant information about the expected position of a particle during the caging regime. However, clearly $\Delta r^\mathrm{CS}$ is a better predictor of the dynamic propensity in all three cases, reaching correlations stronger than 0.8. This is significantly higher than the results shown in Fig. \ref{fig:prop}, despite being based only on a single variable. Evidently, the simple averaging method we used to obtain the cage state indeed manages to eliminate the thermal fluctuations that are present in the initial configuration, without losing significant information about the underlying cage structure. The inherent state performs less well, in particular in the case of harmonic spheres. To shed light on the reason behind this, we measure the radial distribution function $g(r)$ of the initial, inherent, and cage state configurations and compare them in Fig. \ref{fig:cagestates}def. In the harmonic and KA models, both the inherent state and cage state increase the degree of local structure in the system, resulting in higher peaks in $g(r)$ -- and this degree of additional ordering is stronger for the inherent state than for the cage state. This implies that the inherent state quench pushes the system significantly further away from its local structure than the system would normally sample. In contrast, the cage state procedure only takes into account configurations that the system samples during (constrained) thermal fluctuations. Hence, it is perhaps not surprising that the cage state better reflects the expected dynamics of our systems.

We now examine whether knowledge of the cage structure can help us make more accurate predictions of the dynamic propensity even outside the caging regime. To this end, we start with the structural order parameters and linear regression approach described in Refs. \onlinecite{alkemade2022comparing, boattini2021averaging}. In particular, for each particle, we define a set of approximately 1000 structural parameters describing their local structural environment, and use standard ridge regression in order to fit the dynamic propensity as a function of these structural parameters. A full description of these structural order parameters for each model is provided in the SI. Note that this is the same approach as we used for the data in Fig. \ref{fig:prop}. As a basis for calculating the local structural descriptors, we now use either the initial, inherent, or cage state, resulting in three sets of structural descriptors for each particle: $\mathbf{X}^\mathrm{init}$, $\mathbf{X}^\mathrm{IS}$, and $\mathbf{X}^\mathrm{CS}$. We then examine how these different sets of input data influence our ability to predict dynamic propensity. The results are shown in Fig. \ref{fig:correlations}abc for the three model systems. In all cases, at short times the initial state of the system provides the best input for predicting propensities. This is understandable, since only the initial state contains information about the exact particle environments in the limit $t\to 0$. At longer time scales, both the inherent and cage state structures outperform the initial state, with the cage state always outperforming the inherent state. This further supports the observation that the cage state, as determined by our MC approach, represents an excellent approximation for the underlying structure of our system when it comes to understanding its dynamics on time scales around the structural relaxation time.

The obvious next step is to combine this information with knowledge of $\Delta r^\mathrm{cage}$, which we already know provides a strong prediction of dynamics in the caging regime. In Fig. \ref{fig:correlations}def, we plot the correlation between the dynamic propensity and predictions based on linear regression, combining as input $\mathbf{X}^\mathrm{init}$, $\mathbf{X}^\mathrm{cage}$, and $\Delta r^\mathrm{cage}$, for both cage definitions. Overall, we observe a massive improvement in our ability to predict propensity at all time scales beyond the ballistic regime. 

\begin{figure}
\vspace{0pt}
  \raggedright
  a)\\
    \includegraphics[width=\linewidth]{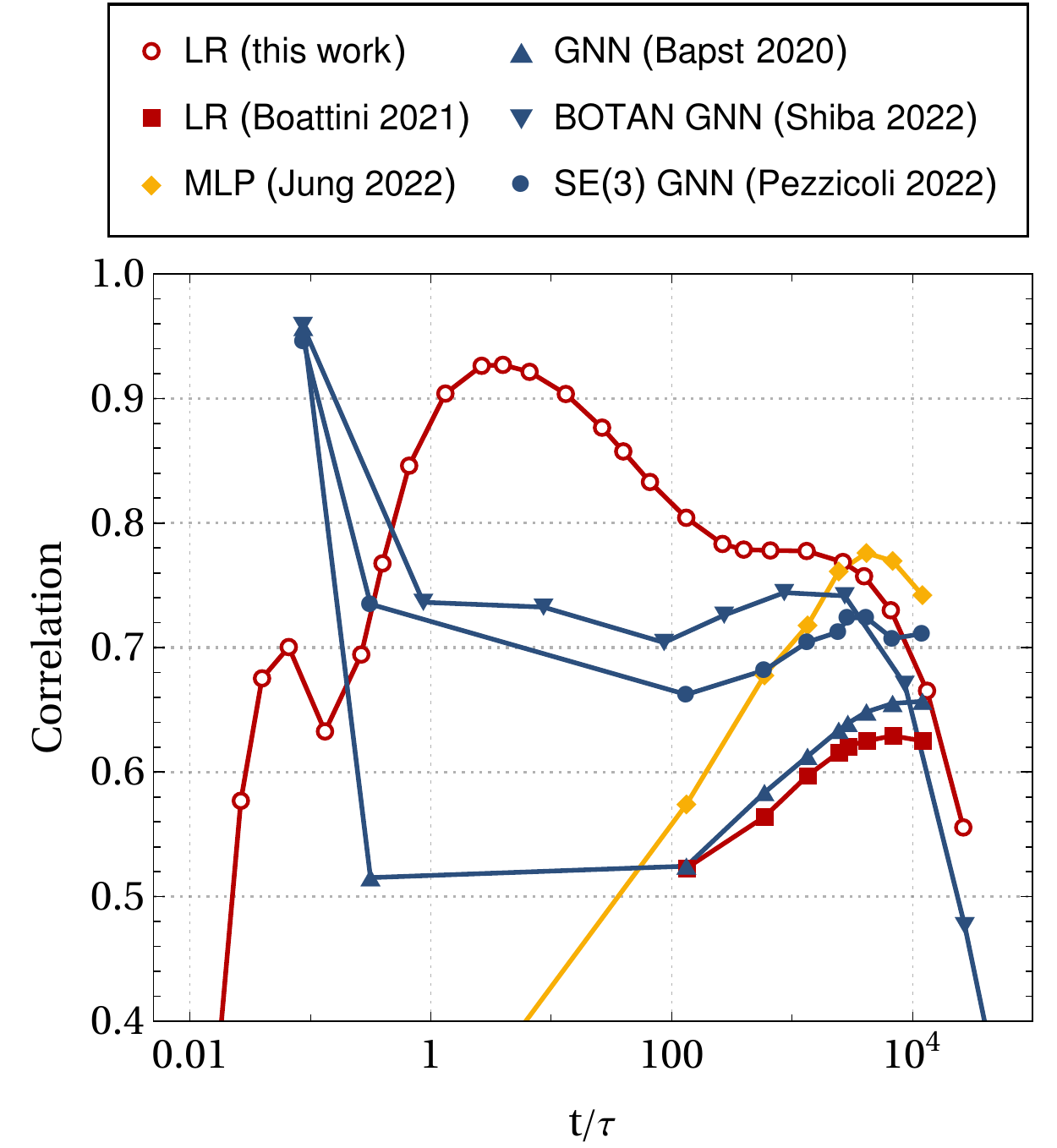}
    b)\\
        \includegraphics[width=\linewidth]{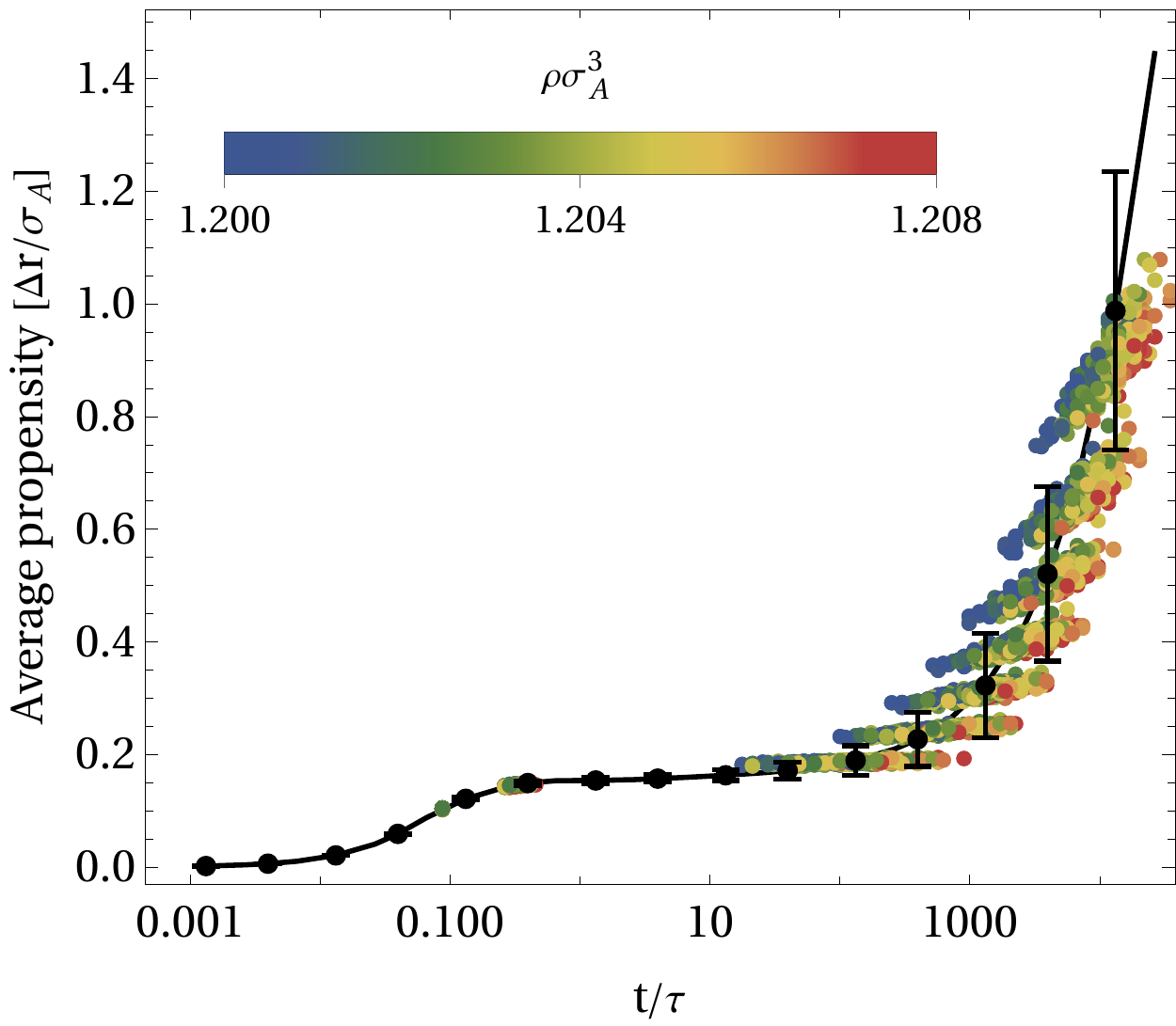}    
\caption{a) Pearson correlation coefficient between the dynamic propensity and the prediction for \textit{A} particles in a KA system at the statepoint used in this paper, made by different models: linear regression (LR) by Boattini \textit{et al.}\cite{boattini2021averaging}, multi-layer perceptrop (MLP) by Jung \textit{et al.} \cite{jung2022predicting}, graph neural network (GNN) by Bapst \textit{et al.}\cite{bapst2020unveiling}, BOnd TArgetting Network (BOTAN) GNN by Shiba \textit{et al.} \cite{shiba2022unraveling} and SE(3)-equivariant GNN by Pezzicoli \textit{et al.}\cite{pezzicoli2022se}. 
b) Average dynamic propensity per snapshot in the dataset of Ref. \onlinecite{bapst2020unveiling}, which was also used in Refs. \onlinecite{boattini2021averaging,pezzicoli2022se,jung2022predicting}. For each initial snapshot in the training set of Ref. \onlinecite{bapst2020unveiling}, we plot the mean propensity in that snapshot at the times reported for that snapshot, and color the point based on the system density. For comparison, we also plot in black the mean propensity taken from our own dataset, where the times and density are the same for each initial configuration. The error bars indicate $\pm 2$ times the standard deviation of the mean propensity at each time.
\label{fig:allKAresults}
 }
\end{figure}

The predictive power of our models using information about the cage state is particularly impressive  compared to past results. Specifically, the state point of the KA mixture we study here has been used in a variety of previous studies where new methodologies were introduced to attempt accurate prediction of the dynamic propensity. In Fig. \ref{fig:allKAresults}, we plot our results from Fig. \ref{fig:correlations}f and compare them to the results of Refs. \onlinecite{bapst2020unveiling, boattini2021averaging, shiba2022unraveling, jung2022predicting, pezzicoli2022se}.  Note that the comparison between different predictions is complicated by significant differences in the datasets we used here and the dataset from the work of Bapst \textit{et al.} \cite{bapst2020unveiling}, which was also used by Boattini, Jung, and Pezzicolo and their respective co-workers\cite{boattini2021averaging,jung2022predicting,pezzicoli2022se}. In particular, for our dataset, we determine dynamics propensities for a set of initial snapshots that are all taken at the same density, and we always measure the dynamic propensity at a fixed set of times. In contrast, the authors of Ref. \onlinecite{bapst2020unveiling} took the initial snapshots from constant-pressure simulations, and hence the configurations vary slightly in density. Additionally, for each initial configuration, they measured the dynamic propensity at different time intervals, based on the decay of the intermediate scattering function for that specific snapshot. In Fig. \ref{fig:allKAresults}b, we illustrate this difference by plotting for each of the initial configurations in the training set of Ref. \onlinecite{bapst2020unveiling} the average propensity as a function of time, and color each point based on the corresponding system density. Hence, in our comparison in Fig. \ref{fig:allKAresults}a, the time for each point based on this dataset is an average time. Clearly, there is a significant correlation between the density of the snapshot and the mean dynamic propensity, and a significant spread in acquisition times. Additionally, the spread in mean propensities is significantly lower than in our own data (shown in black in Fig. \ref{fig:allKAresults}b), due to the grouping of initial configurations based on their structural relaxation instead of time. It should be noted that this grouping is not fully consistent with Eq. \ref{eq:propensity}, and implies that some dynamical information is already included in the input data.

Evidently, including information about the cage state -- captured in our parameters $\mathbf{X}^{CS}$ and $\Delta r^{CS}$ -- allows our linear regression approach to outperform the current state-of-the-art machine learning methods for predicting dynamic propensity over a wide range of time scales. The GNN-based methods do outperform our predictions in the very short-time regime, where the dynamics are likely dominated by the forces that act on the particles in the initial configuration. This may be an indication that these  instantaneous forces cannot be directly recovered from our set of input parameters. In contrast, the three GNN-based approaches plotted (in blue) in Fig. \ref{fig:allKAresults} all have information about the relative positions of particles with respect to their neighbors included in the input graph, and hence might be able to learn the net forces on the particles with high accuracy. At very long time scales, where the motion of particles becomes more diffusive, the results of the different machine learning approaches appear to converge to similar performance -- with the physics-informed approach by Jung \textit{et al.} \cite{jung2022predicting} performing the strongest in this regime. 

\begin{figure}
\vspace{0pt}
    \includegraphics[width=\linewidth]{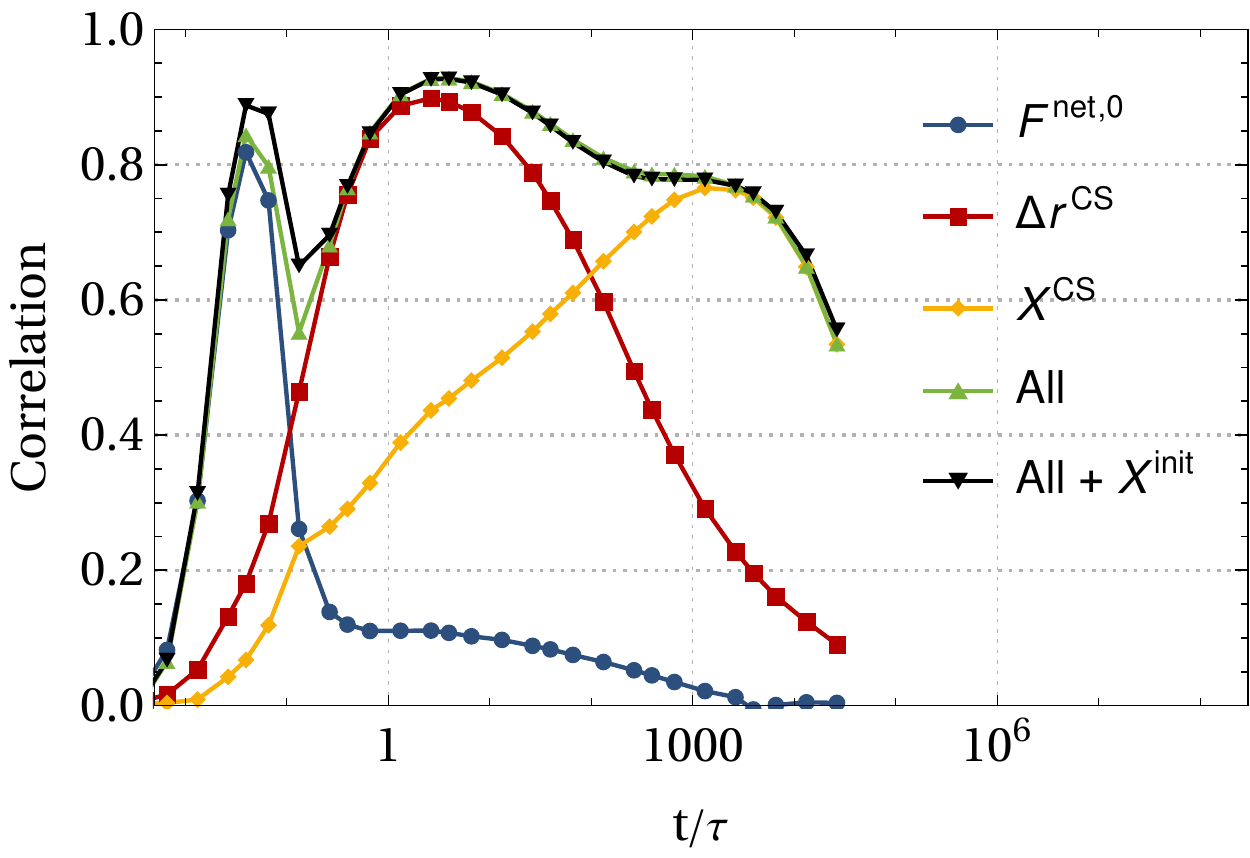}
\caption{Pearson correlation coefficient between  measured and predicted propensity over time, shown for linear regression models trained on different subsets of structural parameters for \textit{A} particles in a KA system. $F^{\text{net},0}$ is the absolute net force that a particle feels at $t=0$, $\Delta r^\text{CS}$ is the absolute difference between the positions of particles in the initial configuration and the cage state and $\mathbf{X}^\text{CS}$ are the structural parameters evaluated for the cage state. Note that the green line (``All'') is the linear regression model trained on the combination of $\{F^{\text{net},0},\Delta r^\text{CS}, \mathbf{X}^\text{CS}\}$, while the black line (``All $+\mathbf{X}^\text{init}$'') also includes the structural parameters of the initial configuration in the structural dataset.  \label{fig:difKAinflunnces}
 }
\end{figure}

To test whether the improved performance of the GNNs at short time scales is indeed due to them directly learning the initial forces, we adapted our method to also include the net force on each particle in the initial configuration. In Fig. \ref{fig:difKAinflunnces} we show the correlation between the dynamic propensity and the initial force in blue, and indeed we see a strong peak at short times. Adding this information into our input data for the full linear regression model, we indeed see a significant improvement at short times (black line). However, the GNN-based methods still achieve a significantly higher accuracy in this regime, and hence appear to be capable of learning more about the short time dynamics than simply the instantaneous forces.

In Fig. \ref{fig:difKAinflunnces} we also break down the our prediction of the dynamic propensity into the different relevant aspects of the input data. Overall, we see that the dynamic propensity at short times is indeed dominated by the forces, at intermediate times by the initial distance to the cage center, and at long times by the structural features of the cage state. While this is in principle not surprising, it is impressive to see that the vast majority of the variation in propensity at short and intermediate time intervals can be explained by just two simple measurements. Moreover, the predictive ability of the cage state structure at long times is impressive as well, and demonstrates that it using the cage state, rather than the initial state to make long-time predictions is an excellent strategy. Note that outside of short times, adding in information on the initial state on top of the cage state is fully irrelevant.

\section{Conclusions}

In short, we have demonstrated that the behavior of the dynamic propensity of glassy fluids can be predicted to high accuracy by using information about the cage state of the initial configuration. This cage state is defined as the set of coordinates describing the average position of the particles when the system is constrained to ensure particles cannot escape their local cages. Combining this information with a simple linear-regression-based algorithm, we can predict dynamic propensities with accuracies that rival or exceed current state-of-the-art machine learning methods at nearly all times. This suggests that the cage state could be a helpful tool for further studying the underlying structure of glassy fluids.

\section*{References}

\section*{Acknowledgements}
The authors would like to thank Marjolein de Jager for many
discussion. L.F. acknowledges funding from NWO for a Vidi grant (Grant No. VI.VIDI.192.102)

\section*{Data Availability Statement}
Raw data associated with the measurements of the dynamical propensity, as well as all figure data, will be published at the following url [???] prior to publication of the paper.

\bibliography{myref}

\end{document}